# NOVELTY AND FORESEEING RESEARCH TRENDS; THE CASE OF ASTROPHYSICS AND ASTRONOMY

Attila Varga

*University of Arizona, School of Sociology, P.O.Box 210027, Tucson, AZ 85721-0027*

ABSTRACT

Metrics based on reference lists of research articles or on keywords have been used to predict citation impact. The concept behind such metrics is that original ideas stem from the reconfiguration of the structure of past knowledge, and therefore atypical combinations in the reference lists, keywords, or classification codes indicate future high impact research. The current paper serves as an introduction to this line of research for astronomers and also addresses some methodological questions of this field of innovation studies. It is still not clear if the choice of particular indexes, such as references to journals, articles, or specific bibliometric classification codes would affect the relationship between atypical combinations and citation impact. To understand more aspects of the innovation process, a new metric has been devised to measure to what extent researchers are able to anticipate the changing combinatorial trends of the future. Results show that the variant of the latter anticipation scores that is based on paper combinations is a good predictor of future citation impact of scholarly works. The study also shows that the effect of tested indexes vary with the aggregation level that was used to construct them. A detailed analysis of combinatorial novelty in the field reveals that certain sub-fields of astronomy and astrophysics have different roles in the reconfiguration in past knowledge.

## 1. INTRODUCTION

It is important for faculties and funding agencies to be able to foresee the future research impact of individuals and research groups when hiring new faculty or awarding research funding (Clauset et al. 2017). A growing body of research is trying to understand how scientific research gains impact. In the past decade, data availability and computing power have allowed scholars to develop sophisticated measures to predict future citation impact of papers, patents, and authors. The problem of predicting research impact can be approached from several directions. First, many researchers are interested in the individual level by focusing on the effect of career stages, past productivity, or institutional affiliation (e.g. Kurtz & Henneken 2015, for a recent review see Clauset et al. 2017). Another line of inquiry aims to foresee the emergence of new and popular technologies and fields of research (Small et al. 2014).

This paper introduces to the reader and improves upon a fertile branch of these efforts that we may call "combinatorial innovation," which aims to theorize and model the process of innovation development and dissemination itself (Schilling & Green 2011, Uzzi et al. 2013, Boyack & Klavans 2014, Kaplan & Vakili 2015, Leahey & Moody 2014, Lee et al. 2015, Youn et al. 2015, Trapido 2015; for an extensive review see: Savino et al. 2017). There are three goals of the current article: 1) to evaluate the measure of combinatorial novelty for citation prediction on several dimensions; 2) to introduce a combinatorial measure for the prognosis of scientific impact based on early anticipation of future trends; 3) to contextualize the relevance of the diversity of combinatorial novelty measures in astronomy and astrophysics.

The main tenet of the paradigm of combinatorial innovation is that novel and original ideas combine past knowledge in a new way (Fleming 2001, Thagard 2012). A measure of combinatorial novelty generally indicates if a combination of knowledge elements is typical or atypical given the past usage pattern of those elements. Despite the diversity of specific research focus and methodologies, all studies found (mostly positive) association between combinatorial novelty and citation impact, and therefore this family of indicators could be utilized to identify trends in science. It must be emphasized that this particular approach to novelty measures the distance/similarity of co-occurrences. This approach can be traced back to early twentieth century, to Schumpeter's (1934) innovation theory, and to Poincare (1910) in the philosophy of science. Poincare stressed that combinations are more innovative and radical if they combine disparate things. The assumption about the relationship between the distance/dissimilarity of ideas and innovativity still has an important role in this paradigm.

"Maximal marginal relevance" is a similar concept in information retrieval (Carbonell & Goldstein 1998, Clarke et al. 2008, recently: An & Huang 2017). It is a search result ranking criteria that assigns higher ranks for relevant records that contain new content than the previously ranked relevant records. A crucial difference however is that while combinatorial novelty evaluates the distance/dissimilarity of pairs of elements, this retrieval concept evaluates only the individual elements in the set. To use the example of Clarke and his coauthors (2008) a search for "jaguar" gives a document that discusses jaguar as a car brand high novelty if the search until that point only retrieved hits about the jaguar as cat. A document that combines the topic of the car and the cat at the first time, is not necessarily marginally novel. If all the information that such a document contains has already been covered by previous search hits about both the car and the animal, the document has low novelty. From the perspective of combinatorial innovation however it is an atypical combination and considered to be highly novel, because it is unusual to combine those elements.

Returning to the literature on combinatorial innovation, the content of documents or the combined elements in research practice is measured typically as bibliometric classification codes assigned by indexing services, keywords, or journals. Although these studies share this basic insight and methodology, they are concerned with different aspects of the innovation process, sample vastly different fields of science and technology, and establish their measure of novelty on various systems of reference list aggregation or keywords. There are several discipline-specific bibliometric classification systems, and other classification systems that journals and citation indexing services assign to publications. One can also use cited journal pairs as building blocks of combinations. The choice of the classification system is sometimes influenced by theoretical considerations. For example, for studies of the citation impact of interdisciplinary research (Lariviere et al. 2015, Yegros-Yegros et al. 2015) and for organizational theorists (Leahey & Moody 2014) atypical combinations are interesting because they span organizations and institutions. Therefore, combined elements must express organizational boundaries. However, the same theoretical construct can be operationalized as a classification system representing subfields (Leahey & Moody 2014), or as Web of Science subject categories representing sub-disciplines (Lariviere et al. 2015, Yegros-Yegros et al. 2015). On several occasions the choice is not justified in detail (Uzzi et al. 2013, Lee at al. 2015).

The paper will test and compare different aggregation levels of reference list combinations to predict high impact papers in astronomy, more specifically examining how the operationalized definition of combinatorial novelty affects its relationship with citation impact. The question of aggregation of scientometric indicators is in fact quite general, and it requires great caution and reflection for students of science of science (Leydesdorff 2001). There are two important dimension of aggregation that will be investigated here: 1) Time: past, present, future; 2) Classification levels: references to papers, to journals, to sub-disciplines. Moreover, since the references to papers (the lowest classification level of ideas in the study) will be investigated herein, "absolute novelty" can also be defined in the context of this paper. While co-citations to journals and subfields are enduring for a long time, the co-citation of particular papers is frequently unprecedented.

The paper will present a diversity of combinatorial novelty indexes, and study their association with citation impact, while aiming to filter out confounding factors. First, choosing a single discipline to study scientometric indicators similar to Kurtz & Henneken (2015) helps to filter out disciplinary effects. For example, Boyack & Klavans (2014) have shown that the results of Uzzi et al. 2013 on the effect of novelty and conventionality of published science papers on citation impact was confounded by disciplinary differences. Studying astronomy in particular helps also to minimize the influence of industry-financed research on novelty seeking behavior of research groups. Evans (2010) argues that biochemistry, which is heavily sponsored by industrial companies, produces research that is less theory driven when the research is sponsored by industry. It is assumed that research in astronomy and astrophysics is primarily concerned with purely scientific questions, therefore the influence of corporate research is negligible. Second, the tested novelty indexes are constructed on three levels of aggregation: references to research documents (articles, or letters), referenced journals, and Web of Science subject categories. The latter is a journal classification system used extensively by researchers studying interdisciplinary science (Porter et al. 2007, Rafols & Meyer 2010). These three levels of aggregation of references can control for the extent of institutionalization from research articles to highly institutionalized sub-disciplines.

A new set of indexes of combinatorial novelty will also be introduced and tested along with the extant measures. If we assume that novel combinations are important, it is straightforward to hypothesize that anticipating the trends in the reconfiguration of science communication early on helps to disseminate new ideas. Aside from combinatorial novelty it will be shown how combinations that are gaining popularity early on or reaching high popularity in the future affect citation impact for publications.

This approach to novelty diverges from the concept introduced above. Instead of quantifying the atypicality of combinations, and attributing novelty to them, in this case the temporal shift of combinatorial activity is measured, and the theoretical assumption is that early trend setting behavior is rewarded by the scientific community. Certainly it is not precluded that trend setting is associated with novelty, but it is more straightforward to interpret the following measures as accurate anticipations in the shift in the focus of a research field. It is assumed that constructing these anticipation measures is another useful test of which level of analysis (articles, journals, and subject categories) is more predictive of future citation gain, and how the dynamics of these levels differ.

Although the study design is explorative and may seem technical, as demonstrated below, it will provide refutations to past theories of how new approaches and discoveries gain footing. The current work not only serves as an introduction to this exciting research agenda for the audience of *ApJS*, it also highlighting how the studied intellectual and institutional factors play out in a specific discipline: astronomy and astrophysics. The final section of the paper will present "maps" of the core journals of astronomy and astrophysics as citation networks, where the behavior of the tested indexes can be investigated in a more concrete fashion. These maps can shed light on how different territories or clusters of publication venues drive trends at different institutional levels in the discipline. This presentation is very important because – given the narrow focus on a single coherent discipline – one can have a contextualized understanding of the meaning of the tested indexes.

## 2. TESTED INDEXES AND DATA

The samples of astronomy and astrophysics bibliographic records were retrieved from Web of Science. All the references are assigned to subject categories based on journal classification. The primary dataset contains all publications (articles and letters) that fall under any physics-related subject category in the past few decades, additionally including the subject category "Multidisciplinary Sciences," which contains all journals that publish articles across the sciences (such as *Science* and *Nature*). To create the relevant sample from the records in the primary dataset, all publications were included that cited at least two journals listed under the subject category "Astronomy and Astrophysics." This sampling procedure is inclusive, instead of sampling only the journals that are categorized as astronomy and astrophysics, several more physics publications were selected for analysis outside of strictly astronomical journals. This sampling approach reflects the general trend toward interdisciplinarity in physics (Sinatra et al. 2015). For 2003 this procedure results in ~17,000 articles. The tests that are presented below are based on the 2003 data, and the Appendix shows key findings for 2007. To calculate past and future frequencies of combinations, and to derive citation impact, a 7-year window was used after and before the sampled year. The above sampling procedure was applied to create both of these windows. I will refer to these three time intervals throughout the paper as present $T_0$, past $T_{-1}$, and future $T_1$.

Relying on references to papers in order to construct an index of combinatorial novelty is unprecedented. One probable reason why it is such a neglected measurement choice is that it gives a large number of data points. Doing so, however, provides a fine-grained and dynamic picture of how communication is configured in science (Small 1978, Leydesdorff 1998). The reference list of a paper is $C = (C_1, C_2 \ldots C_n)$, and a co-citation of two articles in this reference list is denoted as $\{C_i; C_j\}$ ($C_i \in C, C_j \in C$). From $C$ we can derive the aggregated reference lists at the journal and subject category level. Two combined journals and two combined subject categories are denoted as $\{J_i; J_j\}$ and $\{S_i; S_j\}$ respectively. One journal can be assigned to several subject categories, and in these cases all the subject categories were included in the analysis.

To construct an index of combinatorial novelty we must first measure the raw count of combinations. This gives a raw affinity score for the combinations. All these combination frequencies at the three levels were recorded in three time intervals defined above (present $T_0$, past $T_{-1}$, and future $T_1$). Studies of combinatorial innovation create baseline frequencies based on the sum of past and present combinations. The past and future combinatorial frequencies are recorded to calculate the anticipation scores. These raw counts for the "present" are expressed as $F_{T_0}(\{C_i; C_j\})$ for citation pairs, $F_{T_0}(\{J_i; J_j\})$ for journal pairs, and $F_{T_0}(\{S_i; S_j\})$ for subject category pairs. Journals and subject categories exist for a longer time frame, and absolute novelty – when a pair is totally unprecedented in the past ($F_{T_{-1}}(\{J_i; J_j\}) = 0, F_{T_{-1}}(\{S_i; S_j\}) = 0$) – in the sampled time interval is rarer then new paper combinations. In this latter case absolute novelty is defined as:

$$N_{T_0}(\{C_i, C_j\}) = \begin{cases} 0, & F_{T_{-1}}(\{C_i; C_j\}) > 0 \\ 1, & F_{T_{-1}}(\{C_i; C_j\}) = 0 \end{cases}$$

Therefore $N_{T_0}(\{C_i, C_j\})$ is 1 if the articles $C_i$ and $C_j$ have never been cited together before, and zero otherwise. Notice that this index disregards the degree of distance of the combined elements: two entities are either related or not.

The second component of a novelty score is a normalizing weight. This serves to deflate the raw counts of combined pairs if the citations to the individual constituents of the pair are higher. This is sometimes derived from a Monte Carlo algorithm (Uzzi et al. 2013), but is mostly defined as a simple function of the frequency of the individual combined elements and expressing an expected count of the pair (e.g. Leahey & Moody 2014, Lee et al. 2015). Here a simple weight $W = 1/(d_i d_j)$ is defined, where $d_i$ and $d_j$ stands for the number of times the given article $C_i$ and $C_j$, journal $J_i$ and $J_j$, or subject category $SC_i$ and $SC_j$ has been cited individually in the relevant time interval.

Similar to the previously discussed studies, a normalized score expresses the novelty – or more specifically the atypycality – of journal (referred as JR) and subject category pairs (SC). (Table 1 summarizes these scores in a comprehensive way.) The higher these scores, the more "atypical" (or novel) the combination is. For these two scores the recorded time intervals include the past and present. For the raw co-citations to articles a normalized score is also created (CIT) in a similar fashion, which indicates if a combination is less expected given the popularity of combined papers in the past and present. All three scores pertain to combinations; to aggregate them to the level of referencing articles the mean of the scores was calculated. Uzzi and his collaborators (2013) suggest that both novelty and conventionality are important to create a ground-breaking paper. These are two separate dimensions of a publication, and the central tendency of their combination novelty score distribution is used to measure conventionality, while novelty is defined as the tail of this distribution. To include this theory in the test, aside from the mean, the 90[th] percentile of the scores is also investigated below (the score of the most atypical pairs).

NCIT is the index of the percentage of completely new co-cited article pairs $N_{T_0}(\{C_i, C_j\})$. No weighting was applied to this index.

**Table 1**
Novelty indexes Tested in the Study

| Index | Formula | Description |
|---|---|---|
| JR | $F_{T_{-1,0}}(\{J_i; J_j\})W$ [a] | Atypical/novel journal combinations. The number of times journal $J_i$ and $J_j$ have been co-cited at $T_{-1}$ and $T_0$ weighted by the normalization weight $W$. $W$ is based on the past and present frequencies. |
| SC | $F_{T_{-1,0}}(\{S_i; S_j\})W$ [a] | Atypical/novel subject category combinations. The number of times subject category $SC_i$ and $SC_j$ have been co-cited at $T_{-1}$ and $T_0$ weighted by the normalization weight $W$. $W$ is based on the past and present frequencies. |
| CIT | $F_{T_{-1,0}}(\{C_i; C_j\})W$ [a] | Atypical/novel citation combinations. The number of times paper $C_i$ and $C_j$ have been co-cited at $T_{-1}$ and $T_0$ weighted by the normalization weight $W$. $W$ is based on the past and present frequencies. |
| NCIT | $N_{T_0}(\{C_i; C_j\})$ | New combinations. From this binary variable the percentage of articles that have not been co-cited in the past have been calculated. |

**Notes.** [a] $W = 1/(d_i d_j)$ is a normalization factor. $d_i$ stands for the number of times a article, journal, or subject category $i$ has been cited in the given time interval.

Finally six new scores are introduced based on the concept of anticipating future trends in the changing configuration of combinations (Table 2). First there are three scores that contrast the past usage of referencing behavior with their emerging future usage. To define anticipation scores for co-cited articles (ACIT) I simply use the number of times the combination appears in the future, $F_{T_1}(\{J_i; J_j\})$, and normalize it with $W$. $W$ in this case is based on the past and present frequencies. Given how popular the cited papers were separately, it expresses how popular their combination becomes in the future, with a higher index indicating more popular co-citation in the future. The anticipation scores for journals and subject categories are based on percentages. Combinations of journals or subject categories constitute a certain percentage of all combinations in the respective category. The anticipation score for journals (AJR) and subject categories (ASC) is the difference between these future and past percentages. Positive values mean that the combinations become more popular in the future, while negative values indicate that the popularity decreases.

**Table 2**
Anticipation indexes Tested in the Study

| Index | Formula | Description |
| --- | --- | --- |
| ACIT | $F_{T_1}(\{C_i; C_j\})W$ [a] | Anticipation score for cited papers. The simple count of how many times co-cited papers appear together in the future (at $T_1$) normalized with $W$. $W$ in this case is based on the past and present frequencies. |
| AJR | $\dfrac{F_{T_1}(\{J_i; J_j\})}{\sum_{i,j} F_{T_1}(\{J_i; J_j\})} - \dfrac{F_{T_{-1}}(\{J_i; J_j\})}{\sum_{i,j} F_{T_{-1}}(\{J_i; J_j\})}$ | Anticipation score for journals. The difference of the percentages between the past $T_{-1}$ and future $T_1$ frequencies of co-cited journal pairs. |
| ASC | $\dfrac{F_{T_1}(\{S; S_j\})}{\sum_{i,j} F_{T_1}(\{S_i; S_j\})} - \dfrac{F_{T_{-1}}(\{S_i; S_j\})}{\sum_{i,j} F_{T_{-1}}(\{S_i; S_j\})}$ | Anticipation score for subject categories. The difference of the percentages between the past $T_{-1}$ and future $T_1$ frequencies of co-cited subject category pairs. |
| JR (alt.) | $\dfrac{F_{T_0}(\{J_i; J_j\})}{F_{T_{-1}}(\{J_i; J_j\}) + 1}$ | Alternative novelty measure based on the concept of anticipation for journals. The ratio of the frequency of present and past combination of $J_i$ and $J_j$. |
| SC (alt.) | $\dfrac{F_{T_0}(\{S_i; S_j\})}{F_{T_{-1}}(\{S_i; S_j\}) + 1}$ | Alternative novelty measure based on the concept of anticipation for subject categories. The ratio of the frequency of present and past combination of $S_i$ and $S_j$. |
| CIT (alt.) | $\dfrac{F_{T_0}(\{C_i; C_j\})}{F_{T_{-1}}(\{C_i; C_j\}) + 1}$ | Alternative novelty measure based on the concept of anticipation for citations. The ratio of the frequency of present and past combination of $C_i$ and $C_j$. |

**Notes.** [a] $W = 1/(d_i d_j)$ is a normalization factor. $d_i$ stands for the number of times a article, journal, or subject category $i$ has been cited in the given time interval.

The anticipation scores defined above give a sense of what anticipation means in the present context: instead of being an index of typicality, it attempts to grasp the direction of change where a field is shifting. However to calculate the score one has to know the future citation rates for combination at $T_1$. It is not useful for prognosis. Of course this all depends on the time window one choses for $T_1$. If someone wants to evaluate the future impact of recently published articles, one can set this time window to 0. This gives a simplified index also shown in Table 2. This alternative index is the ratio of past and present combination frequencies. One is added to the denominator, so no division with zero occurs. The nominator is always at least one, because one paper in the dataset must make this combination in the present to consider it in our calculation. That index can be interpreted as an indicator if a certain combination takes a fresh momentum. Notice that no weighting is added to this index. It is not necessary because the denominator (past frequencies) already contain information on the popularity of the combined elements. Using the past frequencies instead of the citation impact of the combined elements has a different meaning. While the latter deflates the given score for popular journals or citations, the first only deflates their score if they were already used together. Instead of assuming a "blind variation", the alternative measures encapsulate the field specific usage pattern change.

## 3. RESULTS

Raw citation counts were converted to a binary variable. Several studies employ this strategy, and focus on high impact papers, and use a binary variable that indicates the top 5% of citation impact (Schilling & Green 2011, Uzzi et al. 2013, Lee et al. 2015). Many independent variables remain skewed after log transformation (Figures 1-4). CIT mean, AJR, CIT (alt.) is close to being normally distributed. Several distributions have irregular shape. JR mean is almost bimodal. An interesting observation that we can make just based on the histograms is that most articles cite novel paper combinations. From this perspective novelty-seeking behavior is the dominant strategy. On the average, a paper cites 67% new article combinations (more precisely, these combinations are new in the sense that they did not occur in the past seven years). Both ASC and AJR are also heavier on the positive side, which suggests that the disciplinary diversity increased in astrophysics and astronomy in that time. Indeed the number of new journal pairs increased by 29%, and the new subject category pairs by 39% in the given time period. The growth rate of cited paper combinations was 19%. It is also worth mentioning that the range of ASC is higher than AJR. Journal associations appear to be more conservative than sub-discipline boundaries.

To be able to present the association of citation impact and these other variables concisely and in a similar way for comparison, all independent variables are analyzed by their percentiles. This means that in the following figures and calculations, the probability of a hit paper is plotted along the percentiles of the independent variables, which avoids the analytical drawbacks of these non-normal distributions, and also allows detection of non-linear associations.

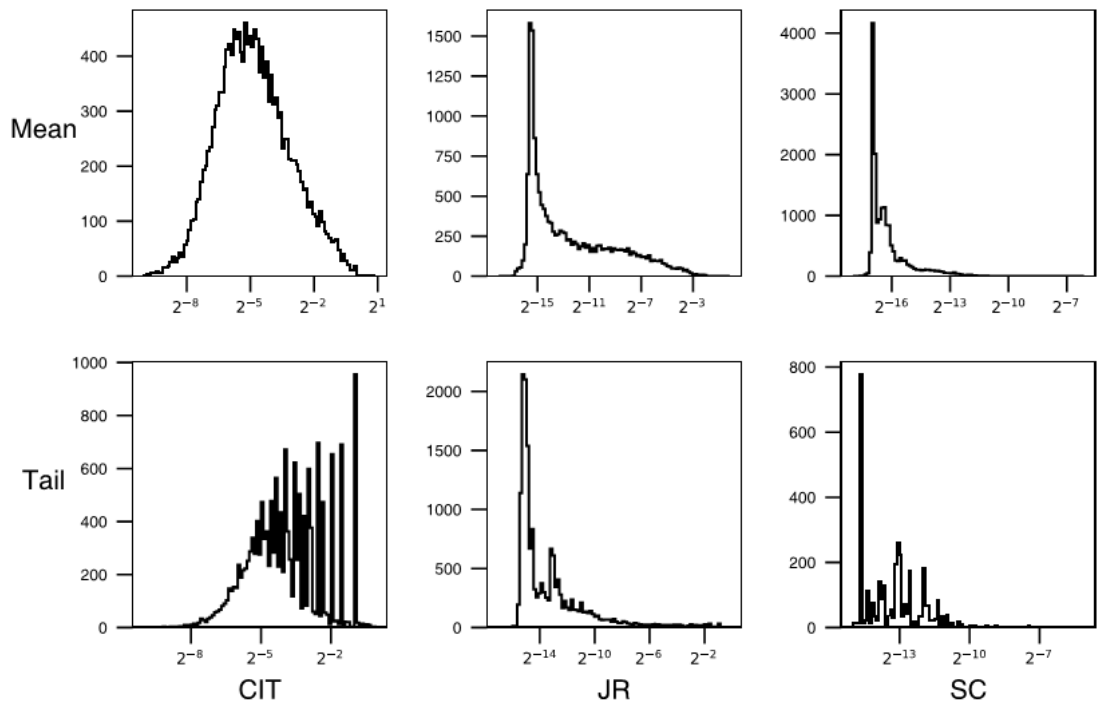

**Figure 1.** Distribution of novelty indexes. The figure shows the distribution of the three novelty scores (by columns), and their respective means and 90$^{th}$ percentiles (by rows).

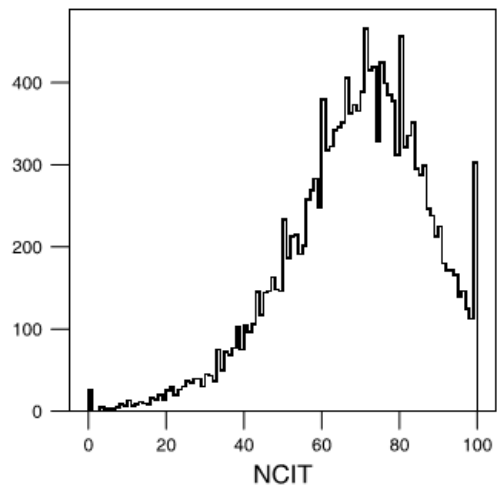

**Figure 2.** Distribution of the percentage of new combinations of research documents.

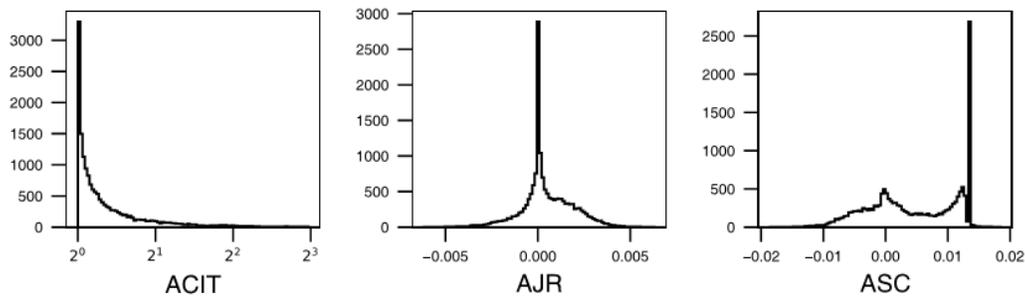

**Figure 3.** Distribution of the mean of anticipation scores. One is added to the ACIT score to fit on the log scale.

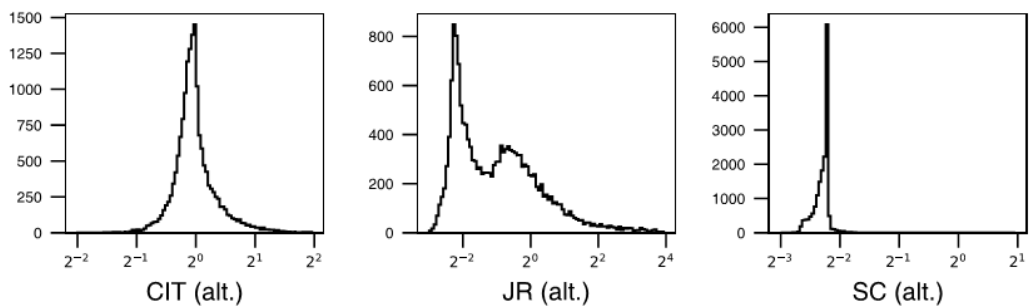

**Figure 4.** Distribution of the mean of "alternative" novelty scores, based on the concept of anticipation.

Surprisingly the novelty measures are not in line with previous findings (Figure 5, Figure A/1). The average of combinatorial novelty decreases citations in the future. This rate is not completely linear for subject category pairs. There are several explanations for these findings. While most studies found a positive relationship between the central tendency of novelty and citation impact (Schilling & Green 2011, Kaplan & Vakili 2015, Leahey & Moody 2014), the study by Uzzi and his coauthors (2013) also found a negative association. Current studies show mixed findings about the success of publications combining sub-disciplines in terms of receiving citations (Lariviere et al. 2015, Yegros-Yegros et al. 2015). Boyack and Klavans (2014) showed that although combinatorial novelty generally brings success to articles, the magnitude of these effects depend on the disciplinary affiliation of the articles. In our case, as we can see later, novelty on the level of subject category pairs varies with the sub-fields of astronomy and astrophysics, and it is rather the citation impact of the sub-fields that explains the citation outcomes of an article. A more general explanation of this finding is that the normalization weight deflates the novelty scores of those combinations that are taking off in the sampled years, which would penalize research that sets up a trend by gaining fresh citations in the sampled years. The negative effect especially seems to be strong for CIT, for which such a novelty index has not been constructed yet.

Taking a look at the effect of the percentage of new article combinations (NCIT, Figure 6, Figure A/2), one can see that it has an inverted U shape: the most successful referencing strategy is to combine new papers, albeit doing so in moderation. This corroborates the observation of Uzzi and his collaborators (2013), and the hypotheses of Schilling and Green

(2011), and Leahey and Moody (2014). All these studies assumed, for various theoretical reasons, that producing atypical combinations has a limit until which the return on citation impact diminishes. These studies used different methodologies, and no any study used NCIT or a similar index as a measure of novelty, but the current findings are in harmony with their expectations.

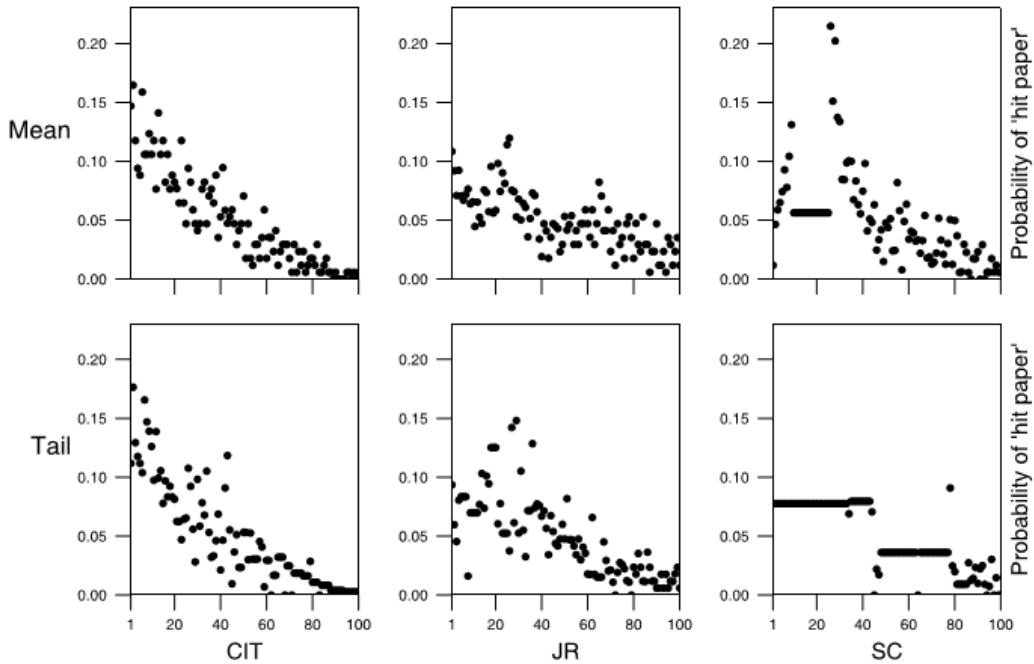

**Figure 5.** "Hit paper" probability by the means and 90$^{th}$ percentiles of the publications' novelty scores (by column). The x-axis is the percentiles of the respective novelty score statistic (mean or 90$^{th}$ percentile, row-wise).

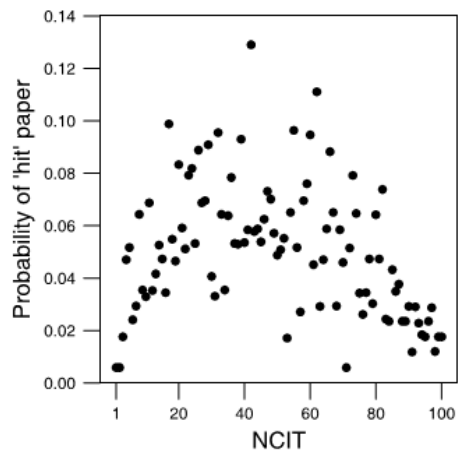

**Figure 6.** "Hit paper" probability by percentage of new combinations.

Turning to anticipation scores (Figures 7 and 8) first one should notice how crucial it is to cite early on papers that will gain popularity later (ACIT). This has a strong association with citation impact. The top 3 percentiles of ACIT have at least ~15% chance to contain high impact articles. CIT (alt.) also have a positive association with citation impact. However, knowing the future co-citation frequencies (like in ACIT for the following 7 years) clearly has a better explanatory power over the future popularity of articles.

The behavior of the indexes that are based on paper citations is markedly different from the ones based on journal and subject category pairs. While the first type can be approximated with a monotonically increasing function (see details later), the latter type has a curvilinear association with citation impact. It is difficult to describe these associations. They are remarkably similar to each other across the two types of measures: AJR is similar to JR (alt.), and ASC is similar to SC (alt.). In the case of journal based measures the highest percentiles of the anticipation scores clearly yield more citations, while the subject category based measures show a fast decrease to the base line 5% at around the 80$^{th}$ percentile. This decrease happens earlier for CIT (alt.). This visual approach also has its limitations. Percentiles are sometimes ambiguous. One can observe on Figures 5, 7 and 8 that the subject category based scores have several percentiles around the 80$^{th}$ percentiles that have the same percentile value.

These results show that while the anticipation of citation trends clearly increases the chance to publish a high impact paper, if the shift involves the crossing of sub-disciplines or research field boundaries the reception of rewards has a more complex mechanism. Later we will see that this phenomenon can be partly explained by the sub-disciplinary structure of astronomy and astrophysics.

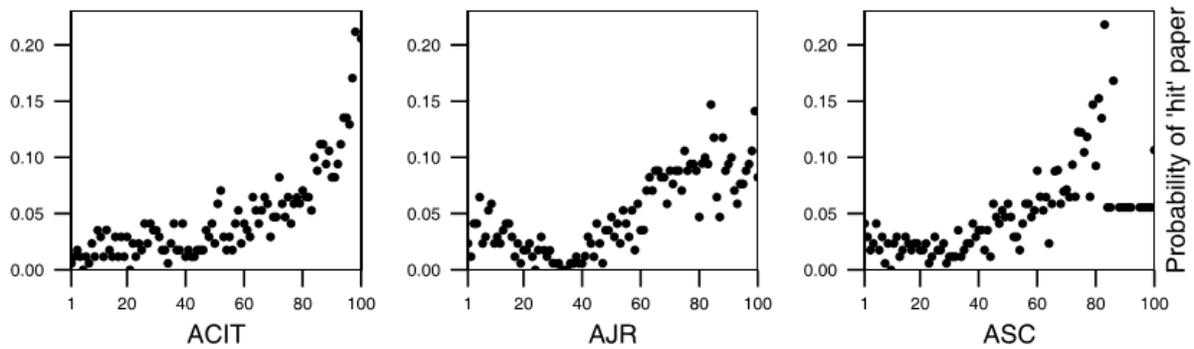

**Figure 7.** "Hit paper" probability by the percentiles of the publications' anticipation scores.

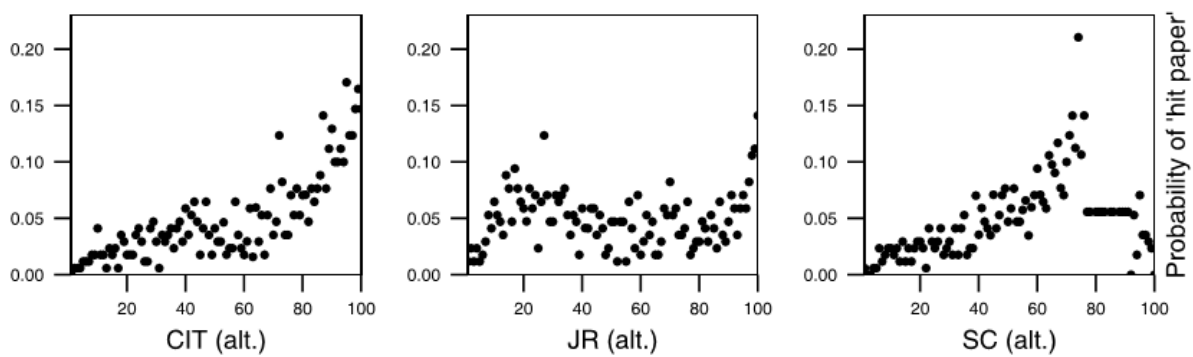

**Figure 8.** "Hit paper" probability by percentiles of alternative definition of novelty scores.

Given these counterintuitive findings about CIT, JR, and SC, and to stay more focused, in the remaining of the article only the anticipation scores will be investigated further.

Table 3, and Figures 9 and 10 show the results of an attempt to fit statistical models on the anticipation/novelty scores. The primary reason to construct statistical models is to compare the magnitudes, and possibly to describe the relationship better. Polynomial logistic regressions have been fitted to the data, using the *R* package *gnm* (Turner & Firth 2015). The Figures (9, 10) plot the logit of the odds of hit papers by the percentiles of the scores. One can see that the models generally fit poorly to the data, especially for the journal and subject category measures. The tests of the residual changes however are all statistically significant (Table 3). The author tried to fit the best functions up to 4-degree polynomials to the data. Using other types of functions in the analysis was difficult because of the frequent non-convergence of the maximum likelihood estimators. Table 3 shows both the bivariate association of the independent variables and the dependent variable, and the hierarchical model where the independent variables entered in the order: citation-, journal-, subject category based metrics. Both the subject category scores and the journal based scores are statistically significant when they enter in the hierarchical models. This suggests that the three levels of aggregation explain different aspects of how to gain citation impact. However we have to be careful to make this conclusion because of the poor fit.

In order to somewhat remedy the problems with the logistic models, mutual information scores are also reported in Table 3. Mutual information was fitted to the binary variables "hit paper", and the percentiles of the scores. This measure does not enforce any particular form of the relationship. According to these results, citation based measures have the best explanatory power. The alternative SC index is better in that regard than the alternative JR index, while AJR is slightly better than ASC. However this ordering is not stable; see the results for 2007 dataset in the Appendix (Table A/1). Although one cannot make a clear judgment about the magnitudes of these relationships, the more linear character makes the paper-based ones more desirable for prognosis.

**Table 3**
Fitted model statistics and mutual information for novelty and anticipation indexes

| Model | | Bivariate model, residual d.[a] | Hierarchical model, residual d.[b] | Mutual Information[c] |
|---|---|---|---|---|
| NULL | $Logit(\hat{Y}) = B_0$ | 6672.3 | 6668.7 | |
| CIT (a.) | $Logit(\hat{Y}) = B_0 + B_1X + B_2X^2$ | 6291.7 | 6288.4 | 0.021 |
| JR (a.) | $Logit(\hat{Y}) = B_0 + B_1X + B_2X^2$ | 6628.6 | 6276.7 | 0.009 |
| SC (a.) | $Logit(\hat{Y}) = B_0 + B_1X + B_2X^2$ | 6459.7 | 6116.2 | 0.017 |
| NULL | $Logit(\hat{Y}) = B_0$ | 6672.3 | 6672.3 | |
| ACIT | $Logit(\hat{Y}) = B_0 + B_1X + B_2X^2$ | 6155.8 | 6155.8 | 0.026 |
| AJR | $Logit(\hat{Y}) = B_0 + B_1X + B_xX^2 + B_3X^3$ | 6430 | 5970.5 | 0.021 |
| ASC | $Logit(\hat{Y}) = B_0 + B_1X + B_xX^2 + B_3X^3$ | 6405.5 | 5917.2 | 0.019 |

Notes. [a] Residual deviance of bivariate logistic regressions. All models are significant at the > 3σ confidence level.
[b] Residual deviance of hierarchical logistic regressions. Variables for the novelty and anticipation scores entered in the order shown in the table. All models are significant at the > 3σ confidence level.
[c] MI stands for mutual information in bits.

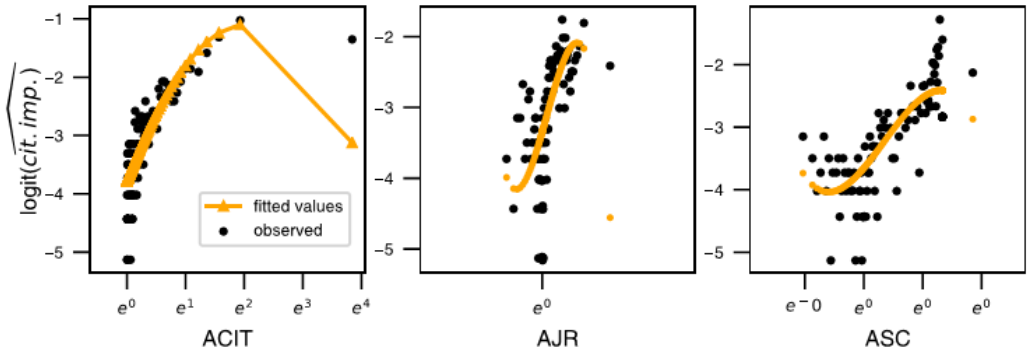

**Figure 9.** Observed and fitted log of the odds of publishing a hit paper as the function of anticipation scores.

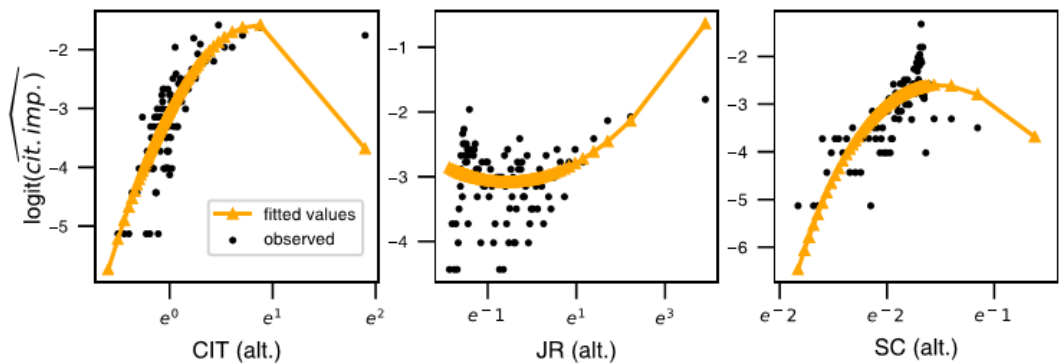

**Figure 10.** Observed and fitted log of the odds of publishing a hit paper as the function of alternative novelty indexes.

Finally we should investigate a little more closely the distribution of these indexes within the discipline. Figures 11-12 shows the citation map of the most important journals that publish articles about astronomy and astrophysics in 2003 and 2007. Tables 4-5 show the indexes for influential journals for these two years. Below, the author provides an interpretation of the visualization, but since this is an exploratory tool, the reader may find more and different approaches to the presented evidence. These maps do not contain all the journals that published relevant articles determined by the sampling method of this paper. The selection criteria for the drawings and the tables was the total citations each journal received in the period under study. Only the top ~70 journals were included for the drawing, and the tables show information for the top 36 journals. The size of the nodes in the figures is proportional to this value. The networks were drawn and manipulated in *Gephi* (Bastian et al. 2009).

      The positions of the journals reflect their relative distance to each other given their propensity to cite one another. The nodes were manually repositioned to make the journal names non-overlapping. A clear division on the map is between physics, and astronomy and astrophysics journals. This reflects the division between theoretical and observational astronomy. The center is occupied by the journals published by the American Astronomical Society, and other generalist astronomy journals. This means that they connect the two fields. On the upper regions one can see the planetary sciences with spectroscopy and physical chemistry. On the physics side one can clearly see a division between certain physics journals that are central for astronomy, and on the other hand specialized nuclear and mathematical physics journals, that are further away from astronomy. One can see that the above described regions changed in this relatively short time interval from 2003 to 2007, most importantly the theoretical and observational side moved closer to each other.

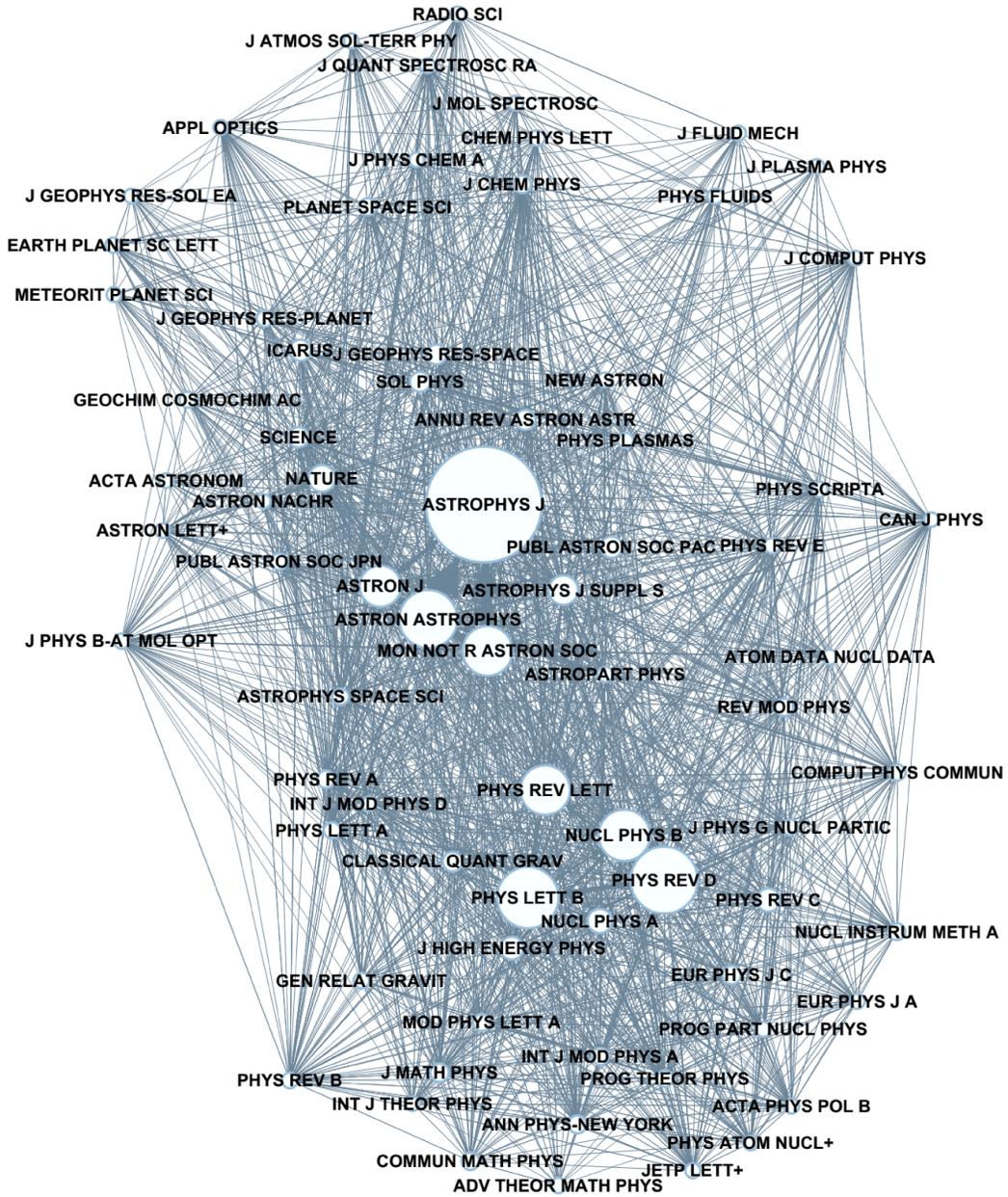

**Figure 11.** Citation networks of core astronomy and astrophysics journals in 2003.

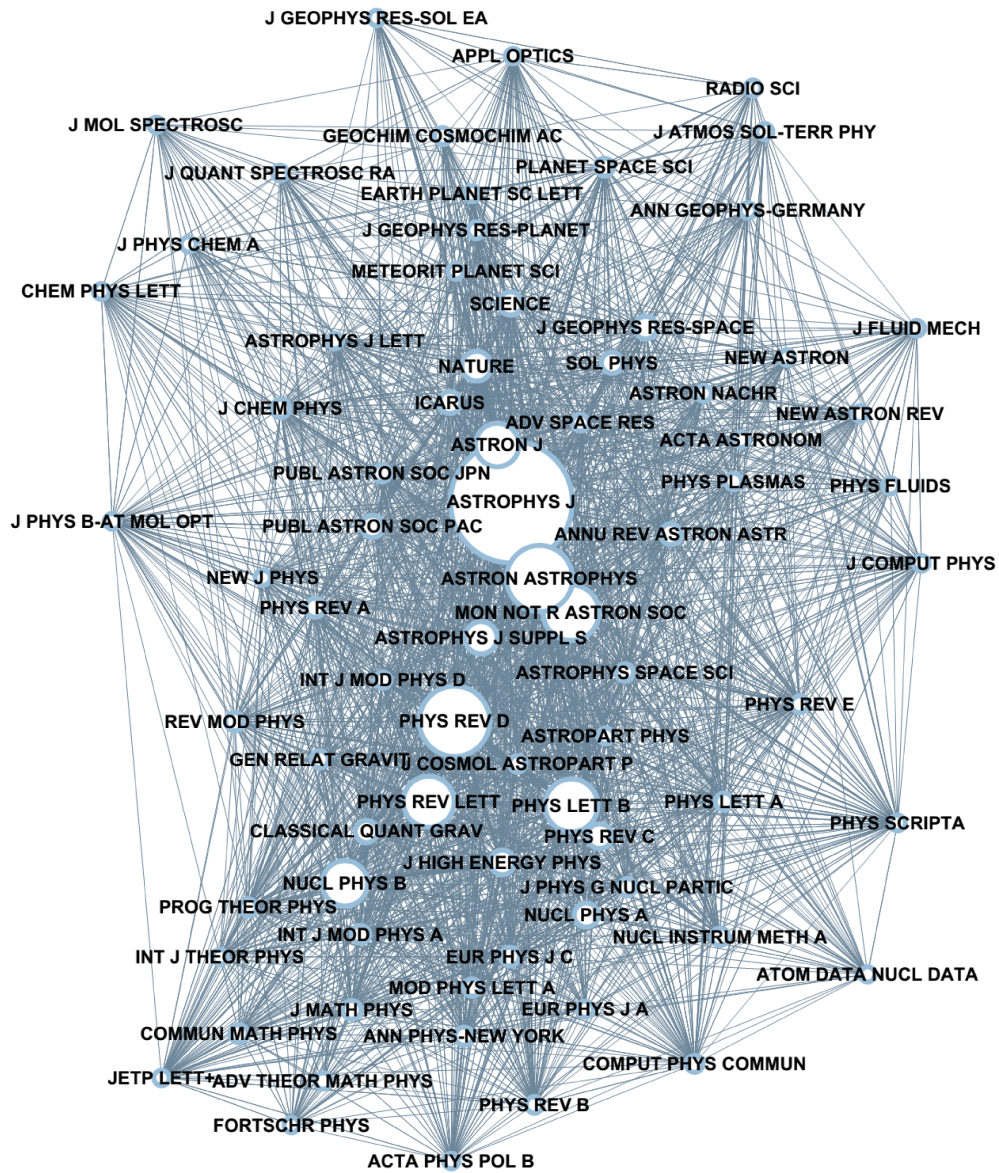

**Figure 12.** Citation networks of core astronomy and astrophysics journals 2007.

**Table 4**
Novelty and anticipation indexes by core journals in 2003

| | CIT (a) | JR (a.) | SC (a.) | ACIT | AJR | ASC |
|---|---|---|---|---|---|---|
| ANN PHYS-NEW YORK | 1.033 | 0.580 | 0.188 | 0.256 | -0.0005 | -0.0023 |
| ANNU REV ASTRON ASTR | 1.014 | 2.629 | 0.218 | 0.103 | 0.0015 | 0.0091 |
| ASTRON ASTROPHYS | 1.019 | 0.780 | 0.213 | 0.519 | 0.0015 | 0.0113 |
| ASTRON J | 1.004 | 0.555 | 0.216 | 0.535 | 0.0013 | 0.0125 |
| ASTROPHYS J | 1.064 | 0.958 | 0.213 | 0.650 | 0.0016 | 0.0110 |
| ASTROPHYS J SUPPL S | 1.153 | 0.668 | 0.209 | 0.725 | 0.0008 | 0.0085 |
| ASTROPHYS SPACE SCI | 1.040 | 0.702 | 0.212 | 0.306 | 0.0013 | 0.0084 |
| CLASSICAL QUANT GRAV | 0.996 | 1.360 | 0.200 | 0.414 | 0.0001 | 0.0005 |
| COMMUN MATH PHYS | 0.926 | 0.731 | 0.191 | 0.203 | -0.0004 | -0.0028 |
| EUR PHYS J C | 0.952 | 0.778 | 0.196 | 0.232 | -0.0007 | -0.0017 |
| GEOCHIM COSMOCHIM AC | 0.947 | 1.041 | 0.261 | 0.129 | 0.0001 | 0.0020 |
| ICARUS | 0.946 | 1.004 | 0.222 | 0.257 | 0.0002 | 0.0057 |
| INT J MOD PHYS A | 0.917 | 0.512 | 0.195 | 0.214 | -0.0006 | -0.0018 |
| J CHEM PHYS | 0.931 | 0.749 | 0.219 | 0.141 | 0.0000 | -0.0003 |
| J GEOPHYS RES-PLANET | 1.009 | 1.779 | 0.240 | 0.311 | 0.0001 | 0.0022 |
| J GEOPHYS RES-SPACE | 0.909 | 0.573 | 0.211 | 0.235 | 0.0000 | 0.0054 |
| J HIGH ENERGY PHYS | 1.250 | 1.008 | 0.199 | 1.157 | -0.0008 | -0.0019 |
| J MATH PHYS | 0.936 | 0.627 | 0.187 | 0.213 | -0.0003 | -0.0018 |
| MOD PHYS LETT A | 1.022 | 0.663 | 0.195 | 0.353 | -0.0004 | -0.0016 |
| MON NOT R ASTRON SOC | 1.029 | 0.875 | 9.362 | 0.524 | 0.0022 | 9.1596 |
| NATURE | 1.325 | 2.402 | 0.209 | 2.322 | 0.0006 | 0.0060 |
| NUCL INSTRUM METH A | 1.010 | 1.150 | 0.182 | 0.189 | -0.0001 | -0.0003 |
| NUCL PHYS A | 0.939 | 0.491 | 0.177 | 0.226 | -0.0005 | -0.0040 |
| NUCL PHYS B | 1.043 | 0.645 | 0.197 | 0.319 | -0.0016 | -0.0029 |
| PHYS LETT A | 0.994 | 0.537 | 0.186 | 0.435 | 0.0000 | -0.0016 |
| PHYS LETT B | 1.114 | 0.762 | 0.193 | 0.561 | -0.0010 | -0.0026 |
| PHYS REV A | 0.920 | 0.611 | 0.173 | 0.154 | 0.0000 | -0.0017 |
| PHYS REV C | 0.938 | 0.532 | 0.175 | 0.320 | -0.0003 | -0.0048 |
| PHYS REV D | 1.100 | 0.951 | 0.202 | 0.842 | -0.0002 | 0.0015 |
| PHYS REV LETT | 1.145 | 0.684 | 0.191 | 0.962 | -0.0002 | -0.0009 |
| PLANET SPACE SCI | 0.929 | 0.964 | 0.230 | 0.142 | 0.0000 | 0.0044 |
| PROG THEOR PHYS | 1.075 | 0.566 | 0.197 | 0.494 | -0.0007 | -0.0010 |
| PUBL ASTRON SOC PAC | 1.049 | 0.800 | 0.214 | 0.889 | 0.0008 | 0.0119 |
| REV MOD PHYS | 1.505 | 0.854 | 0.188 | 2.300 | -0.0001 | -0.0028 |
| SCIENCE | 0.984 | 0.502 | 0.219 | 0.320 | 0.0008 | 0.0050 |
| SOL PHYS | 0.958 | 0.701 | 0.213 | 0.257 | 0.0004 | 0.0098 |

**Table 5**
Novelty and anticipation indexes by core journals in 2007

|  | CIT (a.) | JR (a.) | SC (a.) | ACIT | AJR | ASC |
|---|---|---|---|---|---|---|
| ANN PHYS-NEW YORK | 0.940 | 0.580 | 0.192 | 0.096 | -0.0004 | -0.0054 |
| ANNU REV ASTRON ASTR | 0.973 | 4.038 | 0.253 | 0.150 | 0.0000 | -0.0010 |
| ASTRON ASTROPHYS | 0.975 | 0.746 | 0.219 | 0.347 | 0.0001 | -0.0011 |
| ASTRON J | 0.953 | 0.547 | 0.220 | 0.302 | -0.0003 | -0.0011 |
| ASTROPHYS J | 1.025 | 0.645 | 0.219 | 0.603 | -0.0004 | -0.0010 |
| ASTROPHYS J SUPPL S | 1.148 | 0.584 | 0.217 | 1.427 | -0.0004 | -0.0011 |
| ASTROPHYS SPACE SCI | 1.027 | 0.628 | 0.213 | 0.362 | 0.0000 | -0.0018 |
| CLASSICAL QUANT GRAV | 1.002 | 1.156 | 0.196 | 0.568 | -0.0001 | -0.0056 |
| COMMUN MATH PHYS | 0.970 | 0.917 | 0.207 | 0.202 | -0.0002 | -0.0041 |
| EUR PHYS J C | 0.919 | 0.607 | 0.190 | 0.252 | -0.0012 | -0.0068 |
| GEOCHIM COSMOCHIM AC | 0.981 | 1.140 | 0.275 | 0.160 | 0.0000 | 0.0023 |
| ICARUS | 1.024 | 0.948 | 0.234 | 0.277 | 0.0000 | 0.0001 |
| INT J MOD PHYS A | 1.000 | 0.661 | 0.196 | 0.326 | -0.0008 | -0.0061 |
| J CHEM PHYS | 0.946 | 0.809 | 0.193 | 0.152 | -0.0001 | -0.0005 |
| J GEOPHYS RES-PLANET | 0.935 | 1.030 | 0.245 | 0.290 | 0.0000 | 0.0014 |
| J GEOPHYS RES-SPACE | 0.929 | 0.568 | 0.247 | 0.262 | -0.0002 | 0.0003 |
| J HIGH ENERGY PHYS | 1.122 | 2.092 | 0.191 | 0.804 | -0.0010 | -0.0082 |
| J MATH PHYS | 0.969 | 1.313 | 0.200 | 0.254 | -0.0002 | -0.0042 |
| MOD PHYS LETT A | 1.006 | 0.507 | 0.191 | 0.346 | -0.0008 | -0.0062 |
| MON NOT R ASTRON SOC | 1.014 | 0.711 | 0.219 | 0.550 | 0.0007 | -0.0011 |
| NATURE | 1.096 | 0.881 | 0.228 | 1.006 | -0.0002 | -0.0001 |
| NUCL INSTRUM METH A | 1.028 | 1.285 | 0.223 | 0.282 | -0.0001 | -0.0016 |
| NUCL PHYS A | 0.935 | 0.730 | 0.179 | 0.183 | -0.0002 | -0.0028 |
| NUCL PHYS B | 0.944 | 0.545 | 0.189 | 0.269 | -0.0017 | -0.0085 |
| PHYS LETT A | 0.984 | 0.813 | 0.203 | 0.262 | -0.0001 | -0.0034 |
| PHYS LETT B | 1.051 | 0.753 | 0.189 | 0.571 | -0.0012 | -0.0067 |
| PHYS REV A | 0.983 | 2.058 | 0.197 | 0.275 | 0.0000 | -0.0022 |
| PHYS REV C | 0.950 | 2.042 | 0.178 | 0.314 | 0.0001 | -0.0021 |
| PHYS REV D | 1.055 | 1.737 | 0.194 | 0.790 | -0.0007 | -0.0057 |
| PHYS REV LETT | 1.110 | 1.178 | 0.194 | 0.907 | -0.0004 | -0.0043 |
| PLANET SPACE SCI | 0.984 | 1.145 | 0.254 | 0.262 | 0.0000 | 0.0002 |
| PROG THEOR PHYS | 0.955 | 0.820 | 0.191 | 0.431 | -0.0009 | -0.0063 |
| PUBL ASTRON SOC PAC | 1.032 | 0.868 | 0.222 | 0.385 | -0.0002 | -0.0010 |
| REV MOD PHYS | 0.928 | 0.869 | 0.201 | 0.037 | -0.0002 | -0.0022 |
| SCIENCE | 1.026 | 0.577 | 0.225 | 0.929 | -0.0002 | -0.0001 |
| SOL PHYS | 0.925 | 0.701 | 0.227 | 0.202 | -0.0003 | -0.0007 |

How does this morphology correlate with the indexes? Several observations can be made that support and clarify the earlier observations about our metrics. Let's take our attention first to the citation based metrics. Larger nodes in the center have higher scores, and they set up trends for citation combinations. The *Annual Review of Astronomy and Astrophysics* is interesting in that regard. It is the highest impact factor in the discipline. Reviews have a specific communicative function in a discipline, so it is informative to see the scores this journal receives. While its mean CIT (alt.) score is moderately high, the ACIT score is low. This makes sense, because being a review journal its attention is fixated on the well-established results (combinations) of the past. The journal based metrics have a different distribution over the regions. Astronomy and planetary science show more activity in this dimension in 2003, and the JR (alt.) scores are higher toward the planetary science edge. This difference between the two main regions in that regard is even more pronounced on the longer time scale of AJR. The theoretical physics journals have a conservative, stable journal combination pair structure, while most of the reconfiguration shifts toward the center of astronomy and astrophysics. However this difference is not observable in 2007, when the general physics journals moved closer the "astronomy pole". The *Annual Review of Astronomy and Astrophysics* again has a special behavior on the map, where it is very responsive to the shifts in journal co-citation pairs. The subject category based indexes look similar to the journal based indexes: theoretical physics is more stable than the rest of the map in 2003. One can see the peak of shorter term activity (SC (alt.)) around the planetary sciences and physical chemistry. However, this peak on the longer time span (ASC) shifts to the center of the discipline.

To summarize these observations, the theoretical physics region is more conservative in terms of relying on stable journal and subject category combinations in the long run, while the interdisciplinary activity happens on the observational astronomy side. Again, we can see that the indexes grasp different aspects of the changes in disciplinary practices. While the selection of focus on the literature is happening in the core between theoretical physics and observational astronomy, as one would expect, the interdisciplinary activity is more on the observational side. This "division of labor" to be receptive to institutional changes between the sub-fields of the regions can explain the non-linear association of citation impact and the journal and subject category based indexes (Figures 8-9): more specialized sub-fields and journals with lower citation rates initiate the re-configuration. This is clearly holds for JR (alt.) (compare Figure 9 and Figure 12). The behavior of the *Annual Review of Astronomy and Astrophysics* suggests the hypothesis that the selection of papers happens at a different time scale than the selection of new institutional boundaries represented by typical journal and subject category pairs. While the *Annual Review of Astronomy and Astrophysics* reacts to the latter by being at the $65^{th}$-$90^{th}$ percentiles with its journal and subject category based scores, its citation based scores are around the $45^{th}$-$65^{th}$ percentiles.

## 4. DISCUSSION

The studied indexes of combinatorial novelty have a tangible effect on citation impact in astronomy and astrophysics. The index based on the article level anticipation score (CIT alt.) shown to be a more straightforward tool for prediction of future dissemination of research findings than higher-level predictors, because of its linear association with citation impact. However, the magnitudes of these associations are rather weak. The measures poorly discriminate "hit papers" from the rest. The highest percentage of hit papers in the highest percentiles of the CIT (alt.) indexes are around 17-18%, while the baseline is 5%. This is very far from finding 100% of high impact papers by utilizing such indexes derived from reference lists. At this point it is not possible for scientometrics to fully predict the potential of a paper or a scientist, or to predict the fate of research programs (Clauset et al. 2017). Altogether, metrics that are devised to detect trend changes - referred to here as anticipation scores - have good potential to be used in future studies. The proposed scores have a simple rationale, showing whether a given paper is citing combinations that are gaining popularity at the moment. In other words, this illustrates which combination pairs are getting fresh attention. Future studies may reveal whether the studied associations with citation impact are stable in other fields than astronomy and astrophysics.

This study presents several theoretically relevant findings about how the dissemination of innovation is shaped by institutional factors. It also provides methodological reflections on the better utilization of these indexes. Furthermore, article level scores – which have been neglected by previous research on the topic – have substantial potential. Conventional novelty measures actually have a negative effect on citation impact. Given the contradictory findings about this relationship in the literature, the result is not novel. Creating more new paper combinations (NCIT) is advantageous, but only until a certain threshold, beyond which it becomes a bad strategy. This is not completely a new finding. A previous study by Yegros-Yegros and his collaborators (2015) found similar association between impact and performing interdisciplinary research, although their methodology was different from the one applied here (Porter et al. 2007). Moreover, other studies (Schilling & Green 2011, Leahey & Moody 2014) hypothesized this inverted U-shape, but did not find it with their data and methods. Their reasoning is that spanning boundaries of specialized knowledge is considered to be innovative and beneficial, because one can rely on a greater scope and diversity of information, but it is also risky in several respects. For example, it is hard to master several specializations, and the audience can be resistant or unprepared to consume exported knowledge.

The aggregation level differences of metrics are strong. Not only is the association pattern with citation impact markedly more complex at the journal and subject category level than on the paper level, but the journal citation maps also reveal journal clusters in the discipline to be responsible for trend changes at different levels. In short, theoretical physics is less responsive to institutional changes in the field that are taking off the in investigated time period. In light of this data, astrophysics and astronomy experience some shift in focus in the early 2000s. Heidler (2011) investigates the changing disciplinary relations of astrophysics, and argues that the discovery of the accelerating expansion of the universe in 1998/1999, and that dark energy and dark matter constitute 95% of the universe, induced a growing influence of high energy physics and a growing importance of cosmology, especially in America. Indeed, one can observe by comparing the journal citation maps of 2003 (Figure 12) and 2007 (Figure A/5) that astronomy/astrophysics and the pure physics journals moved closer to each other. Currently the growing importance of exoplanet research and the experimental demonstration of gravitational waves will open up the way for new and exciting research. Studying the processes of how these new findings gain footing is an interesting topic for future scientometric studies.

**Acknowledgements**: I thank the help of Erin Leahey, Philip Rosenfield, Gretchen Stahlman, and the anonymous reviewer for their helpful comments. I am also thankful for the University of Arizona HPC services.

APPENDIX

These figures show the same analysis for the year 2005 (Figure A/1-A/4) and 2007 (Figure A/5-A/8). The curves on these figures are the same as in the main text, and they serve as a reference. The results are very similar across the three years.

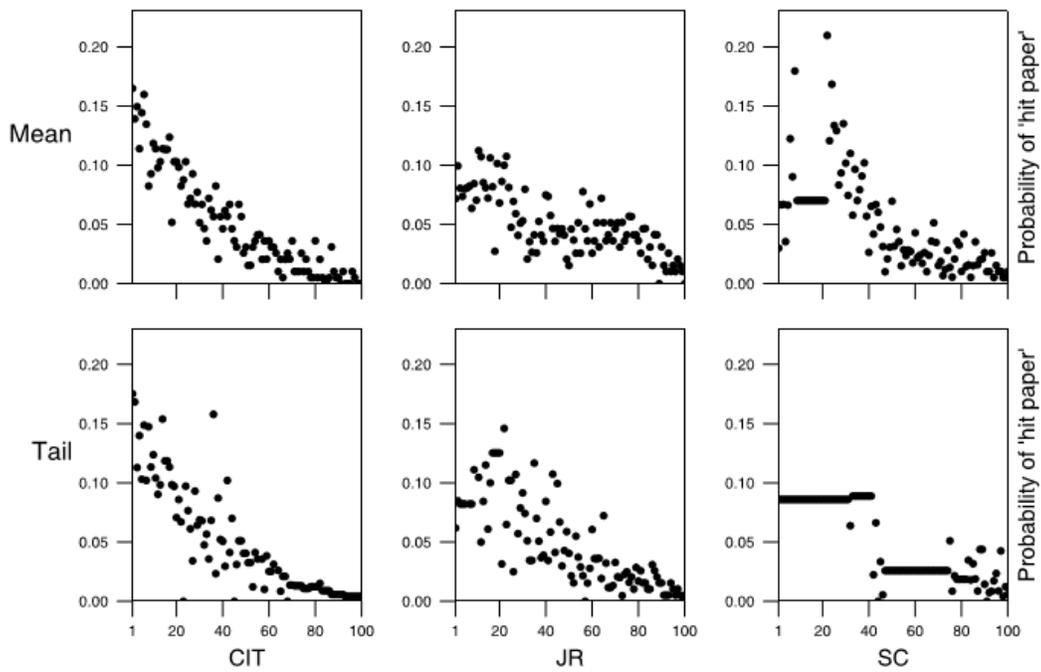

**Figure A/1.** "Hit paper" probability and the means and 90[th] percentiles of the publications' novelty scores (by column) in 2007.

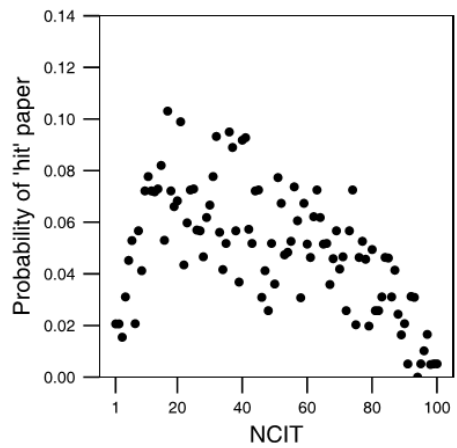

**Figure A/2.** "Hit paper" probability by percentage of new combinations in 2007.

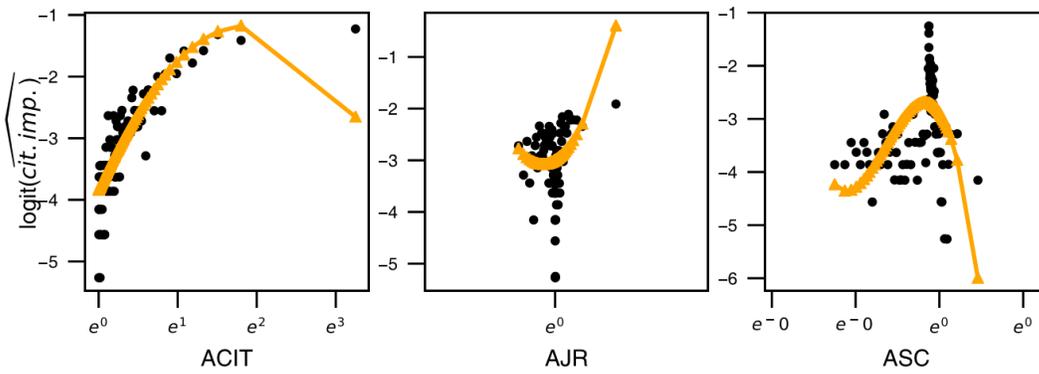

**Figure A/3.** Observed and fitted log of the odds of publishing a hit paper as the function of anticipation scores in 2007.

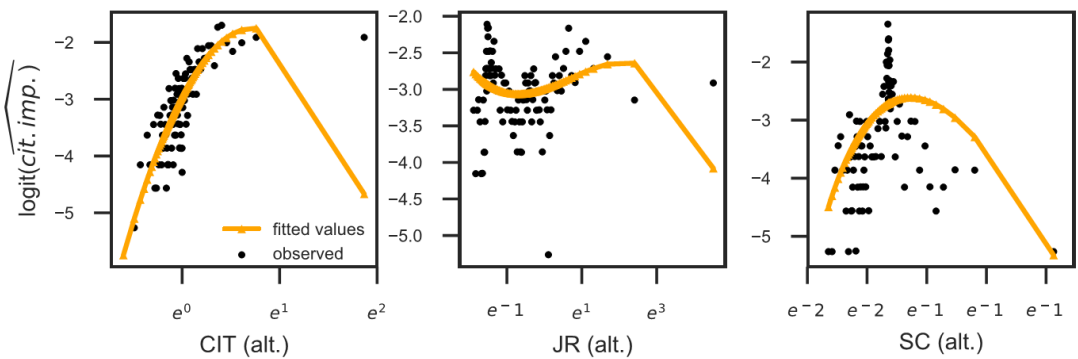

**Figure A/4.** Observed and fitted log of the odds of publishing a hit paper as the function of alternative novelty indexes in 2007.

**Table A/1.**
Fitted model statistics and mutual information for novelty and anticipation indexes for 2007

| Model | | Bivariate model, residual d.[a] | Hierarchical model, residual d.[b] | Mutual Information[c] |
|---|---|---|---|---|
| NULL | $Logit(\hat{Y}) = B_0$ | 7628.7 | 7624.5 | |
| CIT (a.) | $Logit(\hat{Y}) = B_0 + B_1X + B_2X^2$ | 7277.74 | 7272.7 | 0.017 |
| JR (a.) | $Logit(\hat{Y}) = B_0 + B_1X + B_xX^2 + B_3X^3$ | 7619.937 | 7270.1 | 0.007 |
| SC (a.) | $Logit(\hat{Y}) = B_0 + B_1X + B_2X^2$ | 7488.188 | 7122.7 | 0.021 |
| NULL | $Logit(\hat{Y}) = B_0$ | 7628.73 | 7628.73 | |
| ACIT | $Logit(\hat{Y}) = B_0 + B_1X + B_2X^2$ | 7033.885 | 7033.9 | 0.026 |
| AJR | $Logit(\hat{Y}) = B_0 + B_1X + B_2X^2$ | 7590.33 | 7012.6 | 0.011 |
| ASC | $Logit(\hat{Y}) = B_0 + B_1X + B_xX^2 + B_3X^3$ | 7493.774 | 6884.6 | 0.016 |

Notes. [a] Residual deviance of bivariate logistic regressions. All models are significant at the > 3σ confidence level.
[b] Residual deviance of hierarchical logistic regressions. Variables for the novelty and anticipation scores entered in the order shown in the table. All models are significant at the > 3σ confidence level except the addition of JR.
[c] MI stands for mutual information in bits.